\documentclass{sobolev} 

\setpg{1} 

\titl[Continuum and Line Emission of Flares on Red Dwarf
Stars]{Continuum and Line Emission of Flares\\
 on Red Dwarf Stars:\\
\par \bigskip

 Origin of the Blue Continuum Radiation}

\authors{E. S.\,Morchenko}{E. S.\,Morchenko\aff{1}}

\affiliat{
 \item Sternberg Astronomical Institute, Moscow M. V. Lomonosov State University,  Russia}

\email{morchenko@physics.msu.ru} 

\begin{document}

\begin{abstract}

There are two  types of models that explain the appearance of
 the quasi-blackbody radiation during the
 impulsive phase of
stellar flares. Grinin and Sobolev \cite{Grin_77} argue that the
blue component of the optical continuum is formed in ``the
transition layer between the chromosphere and the photosphere.''
Katsova {\it et al.} \cite{Katsova_81} have ``raised'' the source of
white-light continuum up to a dense region in the perturbed
chromosphere. In the present contribution (the main paper is
published in ``Astrophysics'' \cite{Mor_16}), we show that this
statement in \cite{Katsova_81} is erroneous.
\end{abstract}

\section{Introduction}\label{estimation}

Grinin and Sobolev {\cite{Grin_77}} were the first who showed that
the quasi-blackbody spectrum at the flare's maximum brightness is
formed near the photosphere. Heating of the deep layers is due to
the {\it high-energy} proton or/and electron beams with the initial
energy fluxes $F_0\approx10^{12}$ erg\,cm${}^{-2}$s${}^{-1}$ and
$F_0\approx3\cdot10^{11}$ erg\,cm${}^{-2}$s${}^{-1}$, respectively
\cite{Grin_89,Grin_93}.

 Katsova {\it et al.} \cite{Katsova_81} calculated
the first gas dynamic model of the impulsive stellar flares (the
energy flux in the electron beam $F_0=10^{12}$
erg\,cm${}^{-2}$s${}^{-1}$). According to this model, the blue
component of the optical continuum is formed in a {\it chromospheric
condensation}. The condensation is located between a temperature
jump and the front of a downward shock wave (the second-kind
temperature wave \cite{Vol_63}). The physical parameters of this
source of white-light continuum
($N_{\mathrm{H}}\approx2\cdot10^{15}\text{\,cm}^{-3}$,
$T\approx9000\text{\,K}$, and thickness $\Delta{z}\approx10\text{
km}$) lie in the range of the layer parameters in the model by
Grinin and Sobolev {\cite{Grin_77}}
($N_{\mathrm{H}}\sim10^{15}-10^{17}\text{\,cm}^{-3}$,
$T\sim5000-20000\text{\,K}$, and  $\Delta{z}\gtrsim10\text{ km}$).
Here $N_{\mathrm{H}}$ is equal to the sum of the proton and hydrogen
atom concentrations. However, the condensation is formed at a height
of about $1500$ km above the quiescent photosphere of a red dwarf.

The downward {\it non-stationary} shock wave \cite{Katsova_81}
propagates through a {\itshape partially ionized} gas of the red
dwarf chromosphere. The velocities of the flow are subsonic for the
electron component of the plasma, but hypersonic for the ion-atom
component ({\it e.g.,} \cite{Pik_54}). Therefore, both ions and
atoms are heated more intensively than electrons at the shock front
\cite{Katsova_81}. Thus, the region between the temperature jump and
the front of the downward shock wave \cite{Katsova_81} is, in fact,
{\itshape two-temperature} ($T_{ai}\gg{}T_e$). Here $T_{ai}$ is the
ion-atom temperature and $T_e$ is the electron one.
\section{Emission spectrum of a two-temperature layer}
Morchenko {\it et al.} \cite{Mor_15} calculated the emission
spectrum of a {\it two-temperature} ($T_{ai}\neq{}T_e$) homogeneous
plane layer located behind the front of a {\itshape stationary}
plane-parallel radiative shock wave. The layer density lies in the
range
$3\cdot10^{14}\text{\,cm}^{-3}\leq{}N_{\mathrm{H}}\leq3\cdot10^{16}\text{\,cm}^{-3}$;
within the layer  $6\text{\,eV}\leq{}T_{ai}\leq12\text{\,eV}$,
$0.8\text{\,eV}\leq{}T_{e}\leq1.5\text{\,eV}$.

Initially, we have assumed that the Lyman-$\alpha$ optical depth in
the center of the layer, $\tau_{12}^D$, is approximately equal to
$10^7$ (see Eqs. (1) and (53) in \cite{Mor_15}). However, at values
of $N_{\mathrm{H}}\sim10^{16}\text{\,cm}^{-3}$ the layer thickness,
$\mathcal{L}$, is {\itshape small} (as it follows from Eq. (53)
$\tau_{12}^D\propto{N_1}\mathcal{L}$, where $N_1$ is the
concentration of the ground state atoms). Therefore, in
\cite{Mor_15} we consider the transition from the transparent gas to
the gas whose continuous emission is close to the Planck function
under the
 conditions when the value of $\mathcal{L}$ is fixed, and correspondingly $\tau_{12}^D\gtrsim10^7$ (see the first paragraph in
Section 7 in \cite{Mor_15}).

The following elementary processes were taken into account: the
electron impact ionization, excitation, and de-excitation, the
triple recombination, the spontaneous radiative recombination, the
spontaneous transitions between discrete energy levels. We consider
the influence of the layer's radiation (bremsstrahlung and
recombination) on the occupation of atomic levels. It is necessary,
as the flare luminosity is stronger in the optical range than that
of the quiescent atmosphere of the whole star (see, {\it e.g.,} p.
355 in \cite{Grin_77}).

We take into account the scattering of line radiation in the
framework of the Biberman-Holstein approximation \cite{Ivanov_73}.
Since $\tau_{12}^D\gg1$, photons escape the flare plasma in the
distant line wings \cite{Mor_15}. The following {\it asymptotic}
formula is valid for the mean photon escape probability in the case
of the resonance transition:
\begin{equation}\label{Stark}
\theta_{12}\approx\left(\cfrac{\mathcal{B}_{21}\mathcal{E}_{0}}{\Delta\omega_{21}^D}\right)^{3/5}\,\cfrac{1}{({\tau_{12}^D})^{3/5}}\,\,\,\cite{Mor_15}.
\end{equation}
Here ${\mathcal{B}}_{21}$ is the Stark broadening parameter,
$\mathcal{E}_{0}$ is the Holtsmark field strength, and
$\Delta{\omega_{21}^D}$ is the Doppler half-width.

Our calculations \cite{Mor_15} have shown that the Menzel factors of
the {\it two-temperature} layer do not differ from unity at values
of $\tau_{12}^{D}\sim10^7$ and higher. Moreover, the  layer with
$N_{\mathrm{H}}=3\cdot10^{16}\,\text{cm}^{-3}$,
$T_{ai}=10\,\text{eV}$, $T_e=1\,\text{eV}$, and
$\mathcal{L}=10\,\text{km}$ generates the blue continuum radiation
(the optical depth at wavelength
 $\lambda=4170$~{\AA}, $\tau_{4170}$, is approximately equal to 6).

We also {\itshape proposed} that the {\it non-stationary} radiative
cooling of the gas behind the front of a shock wave propagating with
a {\it constant} velocity in the red dwarf {\it chromosphere} toward
the photosphere (``downward'' \cite{Katsova_81}) can produce an
equilibrium region, which is responsible for the quasi-blackbody
radiation at the maximum brightness of stellar flares (the last
sentence in \cite{Mor_15}).

\section{Origin of the blue continuum radiation}
The model \cite{Katsova_81} includes the {\it single-temperature}
($T_{ai}=T_{e}=T$) source of  optical radiation. Let us investigate
the applicability of
 calculation results \cite{Mor_15} for a {\it single-temperature} layer with
$\tau_{12}^D\gtrsim10^7$. It is true that
\begin{equation}\label{Stark_2}
\tau_{12}^{D}\propto\cfrac{1}{\sqrt{\pi}\Delta{\omega_{21}^D}}\propto{T_{ai}}^{-1/2}\,\,\text{(see
Eq. (53) in \cite{Mor_15})}.
\end{equation}
Therefore, the mean photon escape probability $\theta_{12}$ does not
depend on the ion-atom temperature. Thus, at $T_e=T_{ai}=T$
numerical results \cite{Mor_15} remain valid.

Then it is true that the   single-temperature homogeneous plane
layer with parameters from the model by Katsova { \it et al.}
\cite{Katsova_81} is {\itshape transparent} in the optical continuum
(see the lower curve designated to ``I'' in Fig. 2 in
\cite{Mor_15}): $\tau_{4170}\ll1$.

In the paper \cite{Mor_16} we briefly discuss the possibility of
generating  the blue continuum radiation behind the front of a
stationary {\it radiative} shock wave propagating in the red dwarf
chromosphere toward the photosphere. Based on a simple estimate it
is shown that the quasi-blackbody radiation is formed only under
conditions when the gas flows from the discontinuity surface on a
{\it small} distance (approximately five hundred meters). Thus, our
hypothesis \cite{Mor_15} is not confirmed.

Finally, we hold \cite{Mor_16} that the blue continuum radiation
during the impulsive phase of stellar flares originates from the
near-photospheric
 layers \cite{Grin_77,Hou92}.

\end{document}